\documentclass{aa}  
\usepackage{graphicx}
\usepackage{txfonts}
\usepackage{xcolor} % to get colors
\usepackage{graphicx}
\usepackage{lscape}
\usepackage{natbib}
\usepackage{multirow}
\usepackage{txfonts}
\newcommand{\emm}[1]{\ensuremath{#1}}
\newcommand{\emr}[1]{\emm{\mathrm{#1}}}

\newcommand{\unit}[1]{\emr{\,#1}}

\def\EBV{\mbox{\rm{E(B-V)}}}

\newcommand{\Av}{\emm{A_\emr{V}}}

\newcommand{\kms}{\unit{km~s^{-1}}}

\newcommand{\rev}[1]{{\bf  #1}}

\newcommand{\cop}{\mbox{{CO$^+$}}}
\newcommand{\hcop}{\mbox{HCO$^+$}}
\newcommand{\hocp}{\mbox{HOC$^+$}}
\newcommand{\p}{$^+$}
\newcommand{\ps}{s$^{-1}$}
\newcommand{\rmv}{~{\rm dv}}
\newcommand{\cccps}{${\rm cm}^3~{\rm s}^{-1}$}
\newcommand{\HH}{\emr{{\rm H}_2}}
\def\pccc{~{\rm cm}^{-3}} 
 
\def\pcc {~{\rm cm}^{-2}}
\def\fH2{\mbox{f$_\HH$}}
\newcommand{\cotw}{\emr{^{12}CO}}
\newcommand{\coei}{\emr{C^{18}O}}
\renewcommand{\coth}{\emr{^{13}CO}}

\begin{document}

   \title{CO$^+$  as a probe of the origin of CO in diffuse interstellar clouds}

 \author{M. Gerin  \inst{1}
          \and H. Liszt \inst{2}
          }

 \institute{
 LERMA, Observatoire de Paris,  PSL Research University, CNRS, Ecole Normale
  Sup\'erieure, Sorbonne Universit\'e,  F-75005 Paris, France.
  \email{maryvonne.gerin@obspm.fr} %
  \and 
National Radio Astronomy Observatory, 520 Edgemont Road, Charlottesville, VA 22903, USA. 
}
      \date{Received 2020}
% 5 {} token are mandatory
  \abstract
   % context heading
{The chemistry of the diffuse interstellar medium is driven by the combined influences of cosmic rays, 
  ultraviolet (UV) radiation and turbulence. 
 Previously detected at the outer edges of photo-dissociation regions (PDRs) and formed from the reaction of C$^+$ and OH, CO$^+$ is 
 the main chemical precursor of HCO\p\ and CO in a thermal, cosmic-ray and UV-driven chemistry.}
  { To test whether the thermal cosmic-ray and UV-driven chemistry is producing CO in diffuse 
  interstellar molecular gas through the intermediate formation of \cop.}
   {We searched for CO$^+$ absorption with the Atacama Large Millimeter Array (ALMA) towards two quasars with known Galactic foreground absorption from diffuse 
    interstellar gas, J1717-3342 and J1744-3116, targeting the two strongest hyperfine components of 
      the J=2-1 transition near 236 GHz.}
   {We could not detect CO$^+$ but obtained sensitive upper limits toward both targets. The derived upper limits on 
      the CO$^+$ column densities represent about 4 \% of the HCO$^+$ column densities. The corresponding 
        upper limit on the CO$^+$ abundance relative to H$_2$ is  $<1.2 \times 10^{-10}$ . }
   {The non-detection of CO$^+$  confirms that HCO$^+$ is mainly produced in the reaction between oxygen and carbon hydrides 
   CH$_2^+$ or CH$_3^+$ induced by supra-thermal processes while CO$^+$  and HOC$^+$ result from reactions of C$^+$ 
    with OH and H$_2$O. The densities required to form CO molecules at low extinction are consistent with this scheme.}


   \maketitle
%
%-------------------------------------------------------------------

\section{Introduction}

The diffuse interstellar medium (ISM) hosts a rich chemistry where many species reach molecular abundances similar to those
 in denser and darker regions despite the lower densities (a few tens to a few hundred particules per $\pccc$) and the 
relatively unattenuated far ultraviolet (FUV) illumination by the interstellar radiation field 
\citep{SnoMcC06,GerNeu+16,LLPG14,LisPet+14,LisGerBea+18,gerin:19}. Diffuse and translucent interstellar clouds 
%with extinctions ranging from about 0.3 to 3 magnitudes 
\citep{SnoMcC06} have long been used as testbeds of interstellar chemistry \citep{GlaLan75,GlaLan76,BlaDal77,VanBla86,VanBla88}. 
These efforts were successful in explaining the measured 
column densities for \HH\ and small trace molecules such as CH, OH and CN using chemical reactions at temperatures 30 - 80K 
corresponding to kinetic temperatures measured in \HH\ \citep{SavDra+77} although in some cases assuming regions of 
density $\ga 1000 \pccc$  along the line of sight.  However the  observed column densities of CO and the common presence of high
column densities of CH\p\ presented challenges to such models of thermal, cosmic-ray and UV-driven ion-chemistry in diffuse 
molecular gas.

The observed CO column densities are empirically explained by the serendipitous detection of unexpectedly high column densities 
of \hcop\ seen in absorption at 89.2~GHz \citep{lucas:96}. A nearly constant 
abundance of \hcop\ relative to \HH, X(\hcop) = N(\hcop)/N(\HH) = $3\times10^{-9}\pm0.2$dex, is found in comparison 
with OH and CH \citep{gerin:19} whose own fixed abundances relative to \HH\ are observed directly in 
optical/UV absorption \citep{SheRog+08,WesGal+09,WesGal+10}. With such significant column densities, the observed 
\hcop\ ions will recombine with ambient thermal electrons at densities n(H) $\ga 100 \pccc$ to produce the observed 
CO column densities \citep{LisLuc00,VisVan+09}. 

The 1970's-1980's era chemistry is qualitatively correct in predicting a fixed abundance ratio N(\hcop)/N(OH) via 
a dominant chemical chain that proceeds from C\p\ + OH $\rightarrow$ \cop\ + H \citep{GlaLan75,GlaLan76,dagdigian:19} to  
\cop\ + \HH\ $\rightarrow$ \hcop\ + H  and \hcop\ + e$^-$ $\rightarrow$ CO + H. However, the observed N(\hcop)/N(OH) 
$\approx$ 1/30 is roughly thirty times higher than the predicted value through this simple chemical scheme. 
Some \hcop\ can also be produced in this chemistry by the reaction of  C\p\ and \HH O whose abundance is 
28\% that of OH \citep{GerNeu+16,wiesemeyer:16}.  Both reactions\rev{,} \cop\ + \HH\ and C\p\ + \HH O produce equal 
amounts of \hcop\ and the slightly less tightly-bonded isomer \hocp\ that is observed 
with N(\hocp)/N(\hcop) $= 0.015\pm0.003$ \citep{gerin:19}.  This disparity in the
abundances of \hcop\ and \rev{ \hocp\ } is much too large to be explained by the isomerization reaction 
\HH\ + \hocp\ $\rightarrow$ \HH\ + \hcop\ as discussed below.

To reproduce the observed \hcop\  abundance at the low mean densities of diffuse clouds, 
the most successful models include transient production mechanisms that involve dynamical processes 
in magnetized gas such as  low-velocity shocks or turbulent vortices, in which ions and neutral species 
are  partially decoupled \citep{godard:14,lesaffre:20}. 
The resulting slow velocity drift can therefore 
be used as an additional energy source to drive endothermic chemical reactions that could not operate 
otherwise \citep{SheRog+08,VisVan+09}, especially the reaction  
C\p\ + \HH\ $\rightarrow$ CH\p\ + H that is endothermic by 0.36 eV  or about 4300~K.
These models therefore include a specific chemical pathway to CO, which starts  from CH\p, and 
is followed by a series of rapid hydrogen abstraction reactions producing CH$_2^+$ and CH$_3^+$.  
These two ions react with atomic oxygen to produce the required \hcop, and not \hocp.
 In this scheme, \hcop\ is once again the precursor of CO but the predicted CO and \hcop\ 
abundances fit the observations well and the relative rarity of \hocp\ arises very naturally.

Therefore the relative abundances of the three molecular ions \cop, \hcop\ and \hocp\ provide 
key information on the mechanisms at the origin of CO in diffuse gas.  Despite 
the pivotal role played by \cop\ in some versions of the chemistry, its abundance relative to 
\HH\  is much less well constrained than those of \hcop\ and \hocp. \cop\
was first identified in the interstellar medium towards the bright photo dissociation region (PDR) 
M17SW  and the young, high-excitation planetary nebula NGC~7027 \citep{latter:93}. CO$^+$ 
is now routinely observed in dense
PDRs such as the Orion Bar \citep{Goicoechea:17} or MonR2 \citep{trevino:16}. In such regions 
\cop\ is located at the very edge close to the HI/H$_2$ transition, and its abundance 
relative to \hcop\ ranges from one to several tens of percent \citep{fuente:03}.
In  dense PDRs the intense FUV radiation 
can pump H$_2$ into vibrationally excited levels that can reach a significant population \citep{agundez:10}
The internal energy of this vibrationally excited H$_2$ can trigger the endothermic reaction of C\p\ + \HH\ 
forming CH$^+$ as shown by the widespread distribution of CH$^+$ emission in such regions
\citep{goicoechea:19}. Paradoxically, the presence of \cop\ in dense PDRs is a sign that non-thermal
chemical processes are working, while non-thermal processes are invoked in models of diffuse molecular
gas to remove the need to produce \cop\ on the path to CO.

There are no reported detections of \cop\ in the diffuse and translucent interstellar medium: in this paper 
we present  observations with the Atacama Large Millimeter Array (ALMA)  that constrain the abundance of \cop\ there for the first time.  
The observation strategy is presented in section \ref{sec:obs} and the results are discussed in 
section \ref{sec:res} together with the implications for 
the CO chemistry that are emphasized in sections \ref{sec:chem} and \ref{sec:co}.  
The conclusions are summarized in section \ref{sec:conclusion}.

\section{Spectroscopy and observations}
\label{sec:obs}

\subsection{Spectroscopy}

Absorption spectroscopy represents the best method to probe the molecular content of diffuse clouds 
because the molecular excitation is weak at the low densities (few tens to few hundred cm$^{-3}$) and 
pressure (p/k $\approx 3\times 10^3 - 10^4$ K-cm$^{-3}$) of these regions \citep{jenkins:11, gerin:15, goldsmith:18} 
and the level populations are concentrated in the lowest rotational levels.  As a $^2\Sigma$ molecular 
ion, the energy levels of \cop\ are described by three quantum numbers; $N$ the rigid body 
angular momentum quantum number, $S=1/2$ the electron spin angular momentum quantum number and 
$J = N+S = N \pm 1/2$, the total angular momentum quantum number \citep{sastry:81}. 

The \cop\ ground state rotational transitions  at 117.7~GHz and 118.1~GHz are 
close to a strong atmospheric line from molecular oxygen at 118.75~GHz, rendering their observation 
from the ground very difficult. Hence observations of \cop\ have generally targeted excited transitions 
where the sky transmission is much better. We chose to search for the $N=2\rightarrow1$ 
transitions because the frequencies of this first excited transition near 236~GHz are easily 
accessible from ground-based radio observatories and the level population remains significant even 
in diffuse interstellar gas. This $N=2\rightarrow 1$ transition is split into three hyperfine components, 
with the strongest component at 236.06~GHz. Table \ref{table:coplus} lists the three targeted transitions, 
their frequencies, Einstein A-coefficient and the conversion factor between the integrated line opacity and 
the molecule column density $N(\rm{CO}^+)/\int \tau \rmv$ assuming that the energy level population is 
determined by the cosmic microwave background at 2.73~K.

Apart from \cop, the 
setting of the ALMA correlator included spectral windows targeting the $^{13}$CO($2-1$), C$^{18}$O($2-1$) and 
H$_2$CO($3_{0,3}-2_{0,2}$) transitions that were accessible with the same receiver tuning while including a 
dedicated spectral window for continuum phase calibration. Properties of the CO lines are given in 
Table \ref{table:coplus} where
the column density/optical depth conversion is calculated for an excitation temperature of 5\,K that is typical
for this more easily-excited species in diffuse molecular gas \citep{goldsmith:18}.  

%t1
\begin{table}
\caption{Spectroscopic parameters of observed  lines}             % title of Table
\label{table:coplus}      % is used to refer this table in the text
\centering                          % used for centering table
\begin{tabular}{c c c c}        % centered columns (4 columns)
\hline\hline                 % inserts double horizontal lines
Transition \tablefootmark{a}& Frequency & A &  $N(\rm{CO}^+)$/$\int \tau \rmv$\tablefootmark{b}\\    % table heading 
  $N,J$                                           & MHz               & s$^{-1}$  &  cm$^{-2}$ km$^{-1}$s\\
\hline                        % inserts single horizontal line
  2,3/2 - 1 3/2 & 235380.046 & $0.689 \times 10^{-4}$ & $1.01 \times 10^{14}$   \\
  2,3/2 - 1 1/2  & 235789.641 & $3.465 \times 10^{-4}$ & $2.00 \times 10^{13}$ \\
  2,5/2 - 1 3/2 &  236062.553  & $4.172 \times 10^{-4}$ & $1.12 \times 10^{13}$ \\
\hline  
\hline 
 Transition \tablefootmark{a}& Frequency & A &  $N(\rm{^{13}CO})$/$\int \tau \rmv$\tablefootmark{c}\\    % table heading 
  $J$                                  & MHz               & s$^{-1}$  &  cm$^{-2}$ km$^{-1}$s\\
  \hline
2 - 1             &  220398.684            &       $6.076 \times 10^{-7}$                      &$2.42 \times 10^{15}$         \\
\hline
\hline
 Transition \tablefootmark{a}& Frequency & A &  $N(\rm{C^{18}O})$/$\int \tau \rmv$\tablefootmark{c}\\    % table heading 
  $J$                                           & MHz               & s$^{-1}$  &  cm$^{-2}$ km$^{-1}$s\\
  \hline
2 - 1    & 219560.354                            &      $6.012 \times 10^{-7}$                       &   $2.42 \times 10^{15}$       \\
\hline
                               %inserts single line
\end{tabular}
\tablefoot{}
\tablefoottext{a}{The spectroscopic data are extracted from the CDMS database \citep{muller:01,muller:05,endres:16}.}
\tablefoottext{b}{For an excitation temperature of 2.73 K.}
\tablefoottext{c}{For an excitation temperature of 5 K.}
\end{table}

\subsection{Target directions}

The targeted sources are J1717-3342 and J1744-3116, two bright quasars situated at small galactic
latitude behind the Galactic 
bulge that were known to have high intervening column densities of Galactic neutral
gas detected in molecular absorption \citep{gerin:17, liszt:18, RiqBro+18}.  Their positions, 
continuum flux densities S$_\nu$, channel-channel baseline rms line/continuum noise and line
profile integrals and velocity intervals used here as integration intervals for the targeted  \cop\  transitions
are given in Table \ref{table:sources}.  Comparable quantities for the \coth(2-1) and \coei(2-1) lines
are given in Table \ref{table:co}.

%1
  \begin{figure*}
   \centering
   \includegraphics[width=9cm]{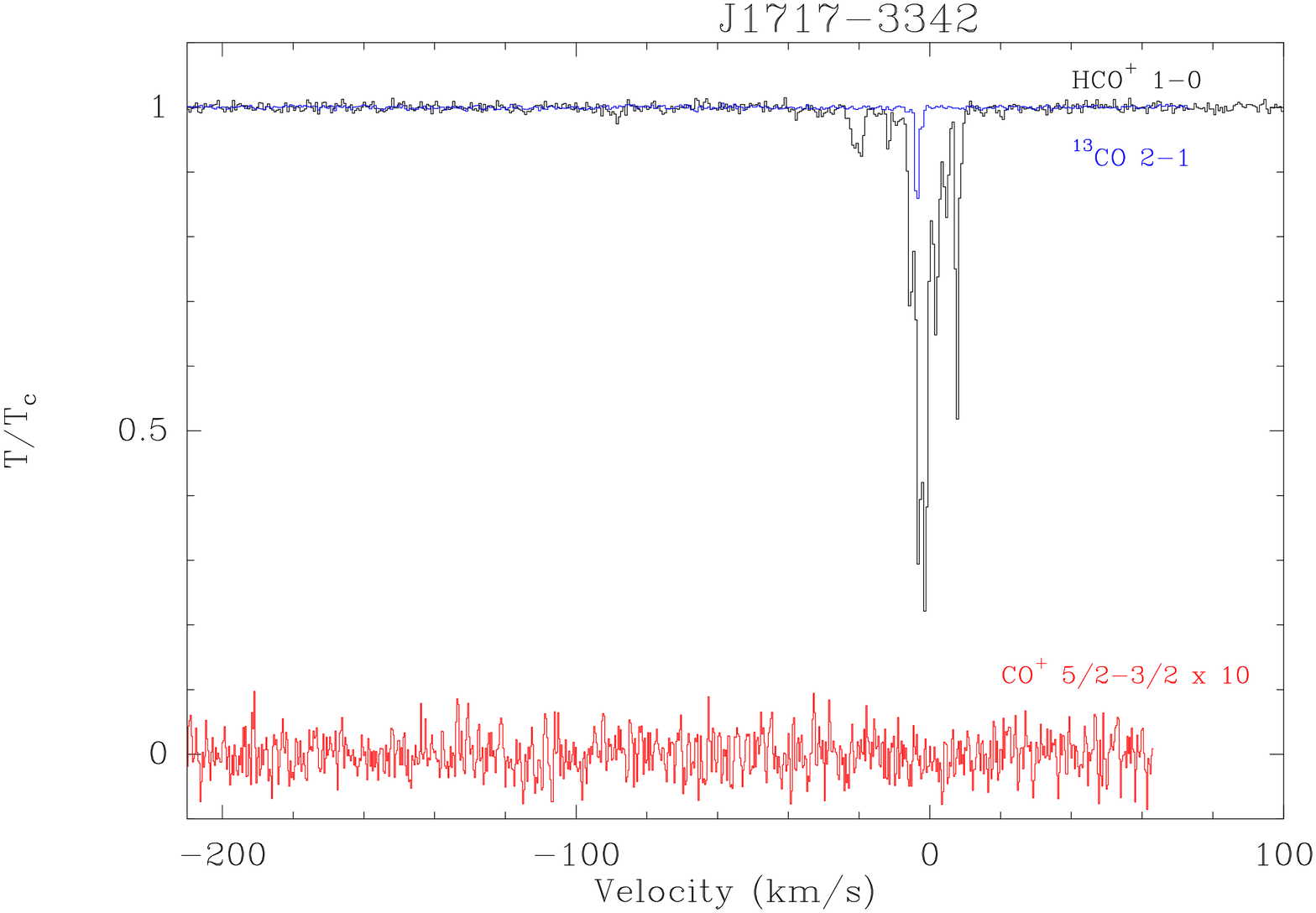}
\includegraphics[width=9cm]{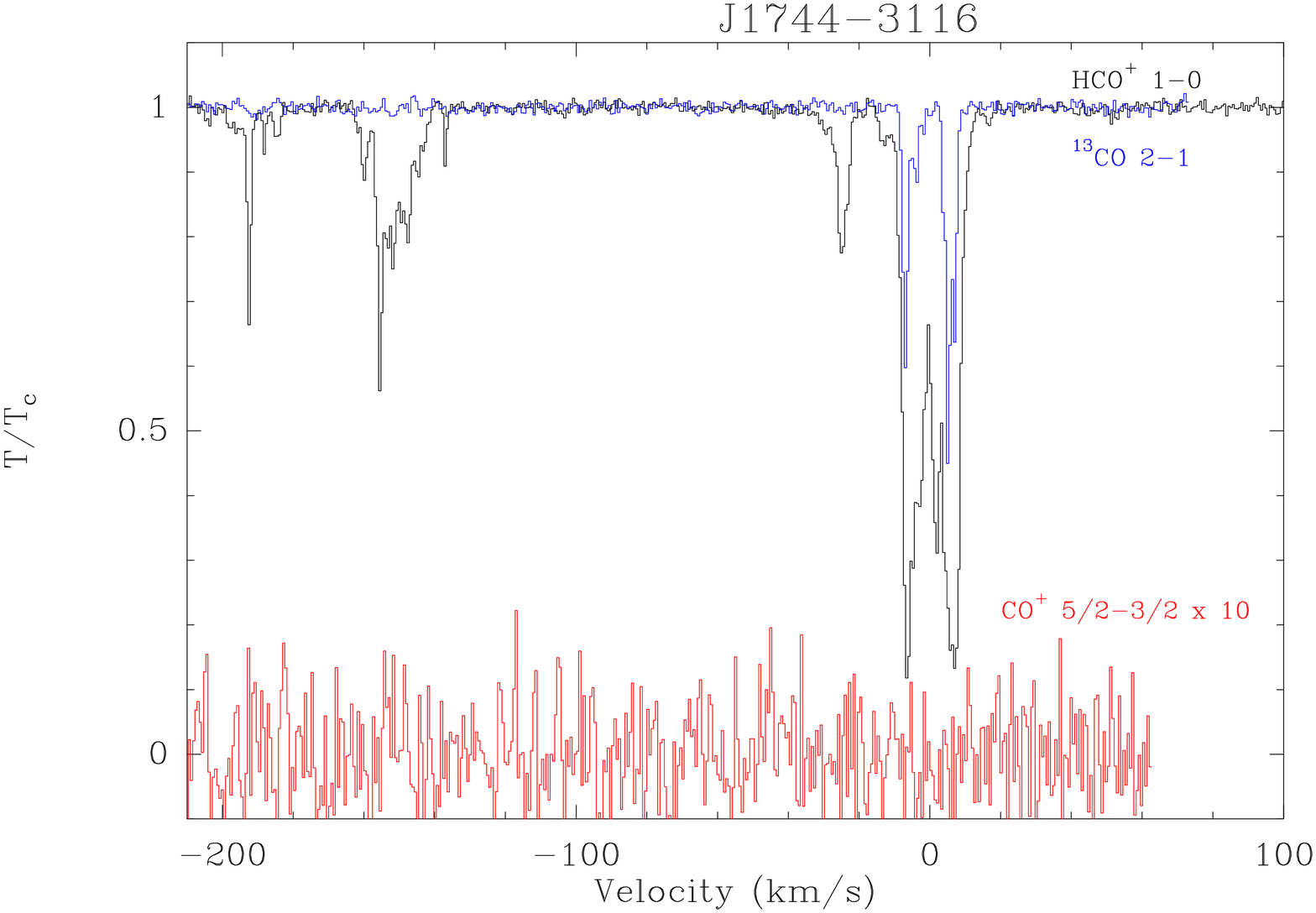}
      \caption{
 Absorption spectra observed towards J1717-3342 (left) and J1744-3116 (right). 
 The spectra have been normalized by the continuum  flux density. 
  The CO$^+$ spectra have been shifted vertically and multiplied by 10. HCO$^+$ is shown 
  in black, CO$^+$ in red and $^{13}$CO(2-1) in blue for comparison. The 
  HCO$^+$ data are taken from \citet{liszt:18} and \citet{gerin:17}.
              }
         \label{fig:coplus}
   \end{figure*}
   
%2
     \begin{figure}
   \centering
  \includegraphics[width=9cm]{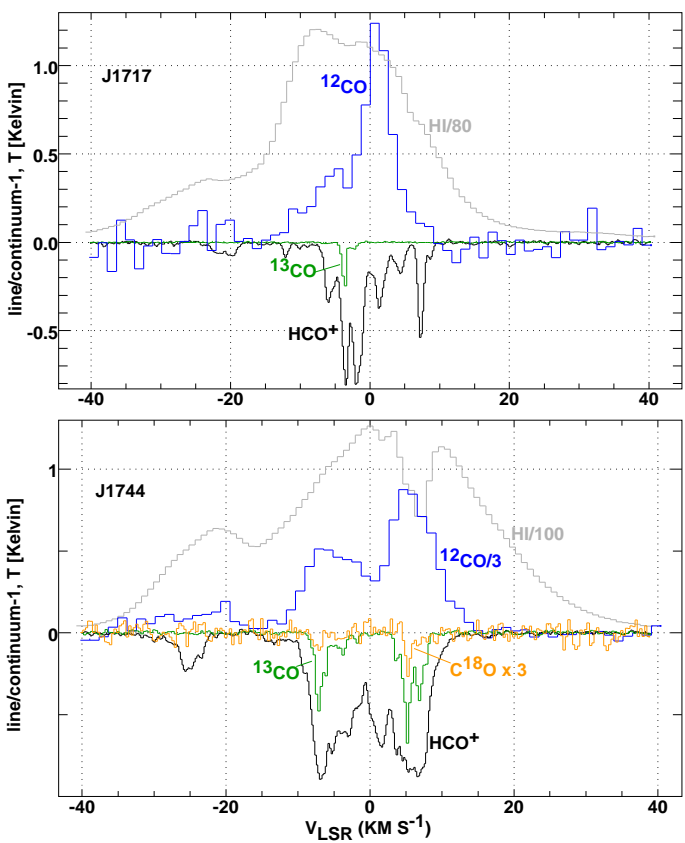}
      \caption{Comparison of the HCO$^+$ (1-0) (black) and $^{13}$CO (2-1) (green) absorption with the 
  $^{12}$CO (blue) and scaled HI (grey) emission. J1717-3342 is shown in the top and 
     J1744-3116 in the bottom. C$^{18}$O(2-1) is also shown in orange for J1744-3116.}
         \label{fig:specemi}
   \end{figure}

\subsection{ALMA observations}

The observations were performed with ALMA band 6 receivers tuned near 236~GHz during ALMA Cycle 7 under the 
project code 2019.1.00120.S. 
The \cop\ lines were observed with a channel spacing of 244 kHz, corresponding to a velocity resolution of 
0.31 kms$^{-1}$.  The channel spectral resolution is twice the channel spacing. As in our earlier observations
toward these sources, the bandpass calibrator was fixed to J1924-2914. Absorption spectra were extracted from
the standard pipeline-processed data products at the peak of the continuum map in each spectral window.
The ALMA spectra of $^{13}$CO,  \cop\  and \hcop\  are displayed in Fig. \ref{fig:coplus}.
The H$_2$CO ($3_{0,3}-2_{0,2}$) line was detected in both directions and will be discussed elsewhere. 

\subsection{Other data discussed here}
\label{subsec:otherdata}

Also shown in Figures \ref{fig:coplus} and \ref{fig:specemi} are 89.2~GHz ALMA \hcop\ J=1-0 absorption line profiles with 0.205 
\kms\ channel spacing and 0.41 \kms\ spectral resolution from our earlier work 
\citep{gerin:17,liszt:18}. Profile integrals are quoted in Table 3, assuming excitation 
in equilibrium with the CMB and using N(\hcop) = $1.12\times 10^{12}\pcc \int \tau \rmv$ as 
before.  Also shown in Fig. 2 are the nearest $\lambda$2.6mm CO J=1-0 emission profiles with an $8\farcm8$ beamsize from 
the work of \cite{BitAlv+97}  and a $\lambda$21cm H I emission profile from \cite{KalBur+05} (FWHM beam of 0.6$^\circ$).
The CO emission profile for J1744-3116 at l,b = 357.875\degr,-1\degr\ is very nearly along the
same sightline; for J1717-3342, the CO profile at 352.75\degr,2.5\degr\ is centered $6\farcm6$ away.

%t2
\begin{table*}
\caption{Summary of \cop\ observations}             
\label{table:sources}      
\centering          
\begin{tabular}{l c c c  c  c c c}     % 8 columns 
\hline\hline   

Source     & $l$  & $b$&LSR Velocity$^a$ & Line & Flux density S$_\nu$& $\sigma_{l/c}$ & $\int \tau \rmv^b$ \\
           & $^o$ &$^o$& km s$^{-1}$                  &      & Jy  &                & km s$^{-1}$     \\
           \hline
J1717-3342 & 352.7333 &2.3911  & -14..10 & CO$^+$~J=5/2-3/2  & 0.780 & 0.0027 &      $<$0.027 \\
           &           &       &         & CO$^+$~J=3/2-1/2  & 0.780 & 0.0034 &      $<$0.033  \\
J1744-3116 & 357.8635 &-0.9968 & -20..20 & CO$^+$~J=5/2-3/2  & 0.234 & 0.0084 &      $<$0.087  \\
           &           &         &       & CO$^+$~J=3/2-1/2  & 0.251 & 0.0107 &      $<$0.111  \\     
\hline                                   %inserts single line
\end{tabular}
\\
$^a$Velocity interval used to integrate the line profile. \\
$^b$Upper limits are 3$\sigma$.\\
\end{table*}

\section{Results and discussion}
\label{sec:res}

\subsection{The sight-lines toward J1717-3342 and J1744-3116}
\label{subsec:other}

Absorption spectra of \hcop, \coth\ and \cop\ toward both sources are shown in Fig. 1, and Fig. 2
presents a comparison of  the absorption spectra of CO and \hcop\ with emission spectra of HI 
and \cotw(1-0) described in section \ref{subsec:otherdata}. Column densities of \hcop, \coth, \coei\ 
and \HH\ (N(\HH) = N(\hcop)/$3\times10^{-9}$) are given in Tables 3 and 4. To compute N(\coth) and 
N(\coei) we assume an excitation temperature of 5 K (Table 1) that is consistent with the 
observed brightness of the \cotw\ emission that is seen toward or near these sources, $1-2.5$ K. 

The diffuse nature of the gas along these lines of sight has previously been discussed by \cite{gerin:17}, 
\cite{liszt:18} and \cite{RiqBro+18}.  High gas column densities accumulate over long paths through the 
Galactic disk at low galactic latitude even without encountering dense clouds. This is indicated in our 
new data by the large ratio of \coth\ and \coei\ integrated optical depths or column densities, 
N(\coth)/N(\coei) = 29$\pm5$ toward J1744-3116 (Table 4). 
If the oxygen  and carbon ratios  have the usual interstellar values 
$^{16}$O/$^{18}$O $\sim 520$ and $^{12}$C/$^{13}$C $\sim 65$ \citep{wilson:94,milam:05,keene:98,giannetti:14}, 
a fully molecular gas would have N(\coth)/N(\coei) = 8 toward J1744-3116. The much larger value we observe 
arises through some combination of selective photodissociation of \coei\ and, more likely, fractionation of 
\coth\ through endothermic carbon isotope exchange when the majority of the free gas phase carbon is in 
C\p\ \citep{Lis17CO}. Our value for N(\coth) toward J1744-3116 in Table 4 agrees with that given by 
\cite{RiqBro+18} based on J=1-0 absorption, N(\coth) $= 6.3 \times 10^{15}\pcc$.

In our earlier work we also derived carbon isotope ratios H$^{12}$\cop /H$^{13}$\cop\ = $58\pm9$ 
and $64\pm4$ in the low-velocity gas toward J1717-3342 and J1744-3116, respectively. These values are 
typical for Galactic disk gas near the Sun and quite different from what is observed inside the 
Galactic bulge.  Thus although the sight-lines employed here cross the Galactic bulge, the 
low velocity gas that is analyzed in CO and \cop\ resides relative nearby in the disk and will be
discussed under the assumption that local conditions are applicable.

%t3
\begin{table*}
\caption{\hcop\ and \cop\ column densities} % title of Table
\label{table:column}      % is used to refer this table in the text
\centering                          % used for centering table
\begin{tabular}{c c c c c} 
\hline\hline   
Source     & N(\hcop) & N(\cop)  &N(\cop)/N(\hcop) & N(CO$^+$)/N(\HH)$^a$  \\
           & $10^{12}$ cm$^{-2}$ & $10^{12}$ cm$^{-2}$ &                     \\
           \hline
J1717-3342 & $7.49\pm 0.03$        & $< 0.30  $  &  $<$ 0.040   & $ < 1.2 \times 10^{-10} $ \\
J1744-3416 & $23.35 \pm0.06$      &  $< 0.98  $  &  $<$0.042    & $ <1.3 \times 10^{-10} $   \\
\hline                            
\end{tabular}
\\
$^a$N(\HH) = N(\hcop)/$3\times 10^{-9}$.
\end{table*}

Overall the CO(1-0) emission is spread over the same velocity range as the \hcop\ absorption. Towards 
J1744-3116, there is a fair correspondence of the main CO emission peaks with the \hcop\ absorption 
features but the \hcop\ absorption profiles are more complex. The positive velocity component towards 
J1744-3116 is detected in $^{13}$CO(2-1) and C$^{18}$O(2-1)  absorption and is associated with a self-absorption 
feature in the H I spectrum that was discussed in \cite{liszt:18}. The presence of the three CO isotopologues 
indicates that this velocity component has probably the largest extinction and mean density and 
could be associated with translucent rather than diffuse gas. There is  no self-absorption 
in the HI profile coincident with the negative velocity component, but the main $^{13}$CO(2-1) 
absorption is associated with a faint C$^{18}$O(2-1) feature.

The CO(1-0) emission differs more from the \hcop\ absorption towards J1717-3342 where the direction of the
emission spectrum is 6.6\arcmin\ away. Of the 8 main HCO$^+$ 
absorption features, only one is detected in $^{13}$CO(2-1) and none in C$^{18}$O(2-1) showing that the gas 
is more diffuse along this line of sight. The CO emission is weaker than towards J1744-3116, barely reaching 1\,K. 
As discussed by \citet{liszt:10}, in diffuse gas the CO(1-0) emission is highly variable as it traces the regions 
where CO reaches column densities above N(CO) $=10^{15}\pcc$ that are high enough to produce detectable emission. 
This transition occurs over a small range of H$_2$ column densities and depends on the local physical conditions
but CO emission at the level of 1 K-\kms\ integrated intensity is not  observed in regions where
the molecular fraction of H-nuclei in \HH\ is much below 0.6.

The comparison of the emission and absorption profiles therefore demonstrates that the two observed lines of 
sight encounter matter with variable physical conditions, ranging from diffuse to translucent gas. The upper 
limits for the CO$^+$ integrated opacities shown in Table \ref{table:sources}  provide constraints on the CO$^+$ 
column densities and abundances for the whole range of physical conditions. For the diffuse molecular gas 
sampled in absorption the molecule excitation is 
dominated by the cosmic microwave background because of the low gas densities. Therefore, the CO$^+$ column densities have 
been derived assuming an excitation temperature of 2.73~K. They are reported in Table 3 together 
with the HCO$^+$ column densities in the same velocity intervals. The achieved upper limit on CO$^+$ column 
densities leads to a very low value for the N(CO$^+$)/N(HCO$^+$) ratio, of about 0.04. The corresponding upper 
limit on the CO$^+$ abundance relative to H$_2$ is X(CO$^+$) $< 1.2 \times10^{-10}$, using the standard value 
for the HCO$^+$ abundance relative to H$_2$ in the diffuse molecular gas, 
$3\times 10^{-9}$ \citep{lucas:96,gerin:19}. These values are a factor a few times higher than those 
established  for HOC$^+$, N(HOC$^+$)/N(HCO$^+$) $= 0.015$ and  X(HOC$^+$) $= 0.45 \times10^{-11}$ 
\citep{gerin:19}.

\subsection{\cop\ as a source of \hcop}

Despite the excellent sensitivity with achieved 
fractional rms flux density levels $\Delta{\rm S}_\nu/{\rm S}_\nu$ of 0.22\% and 0.83\% for 
J1717-3342 and J1744-3116 respectively, \cop\ was not detected towards these sources. The derived 
$3\sigma$ upper limits N(\cop)/N(\hcop) $\la 0.04$ and N(\cop)/N(\HH) $< 1.2-1.3\times10^{-10}$
are similar for both sources in Table \ref{table:sources} because J1717-3342 with smaller
intervening column density is much brighter in the continuum.
The achieved upper limit on the relative abundance is somewhat above the values seen in dense PDRs 
where emission lines from CO$^+$ are detected and X(\cop) reaches a few times 10$^{-11}$. 
They are below the relative abundances seen in the exceptional case of the young planetary nebula 
NGC~7027 where N(\cop)/N(\HH) $= 5\times 10^{-9}$ \citep{fuente:03}.

It is straightforward to show that the relative abundance of \cop\ is too small to provide a major 
source of the observed \hcop\ in diffuse molecular gas, even independent of the \hcop abundance.  
In chemical terms, the reaction \cop\ + \HH\ $\rightarrow$ \hcop\ + H forming \hcop\ can be compared 
with the rate at which \hcop\ recombines with electrons with rates as given in Table \ref{table:chem}.

Equating formation and recombination of \hcop\ , neglecting photodissociation that is
four orders of magnitude slower according to the Kinetic Database for Astrochemistry (KIDA)  \citep{Kida2014}

$$ k_2~n(\cop)n(\HH) = \alpha(T)~n(\hcop)n(e) $$ 

and rearranging, we have

$$ n(\cop)/n(\hcop) = (\alpha(T)/k_2) n(e)/n(\HH) $$

or equivalently

$$ n(\cop)/n(\hcop) = (2/\fH2)~(\alpha(T)/k_2)~x(e)  $$

where \fH2\ is the fraction of H-nuclei in \HH, $x(e)$ is the electron fraction, $n(\HH)$ is the density of \HH\ molecules,
 and $n({\rm H})$ is the total density of H-nuclei. 
Taking $x(e)= 2\times10^{-4}$ corresponding to the observed free gas-phase carbon 
abundance $n({\rm C}^+)/n({\rm H}) = 1.6\times10^{-4}$ \citep{SofLau+04} with a small added 
contribution from cosmic ray ionization of atomic hydrogen in largely molecular gas, the
required amount of \cop\ needed to restore the \hcop\ lost to recombination is

$$ n(\cop)/n(\hcop) = (0.6/\fH2) \times (40/T)^{0.69}  $$

This is some 25 times higher than our upper limit if \fH2\ = 0.6 and T = 40 K.  We noted 
above that CO emission even at the low levels shown in Fig. 2 does not arise in regions 
of very small molecular fraction and temperatures high enough to trigger the formation of CH$^+$, such regions  are only 
characteristic of regions of strong turbulent energy dissipation. 

Alternatively, equating the \hcop\ formation and recombination rates once again but 
setting $n(\hcop)/n(\HH) = 3\times 10^{-9}$ and $x(e) = 2\times10^{-4}$ 
= $2\times10^{-4}(2n(\HH)/\fH2)$, one derives for the required relative
abundance of \cop\ 

$$ X(\cop) = (1.8\times10^{-9}/\fH2) \times (40/T)^{0.69}. $$

This is comparable to the \cop\ relative abundance seen in NGC~7027
($\sim 5\times 10^{-9}$, \citet{fuente:03}) and is at least 15 times 
above the upper limit achieved here, $1.2 \times 10^{-10}$, if T = 40 K
(Table 2).
In this simplified analysis, reducing 
X(\cop) below our upper limit would require temperatures of about 2000 K 
that are more nearly characteristic of the turbulence-driven chemistry, which
can only be present in small and intermittent regions of space and time, where 
other thermal and non-chemical processes are operating. \cop\ is responsible for 
producing at most a few percent of the \hcop\ that recombines with ambient thermal
electrons to form  CO in diffuse molecular gas.

 \section{The thermal chemistry of \cop, \hocp\ and \hcop}
\label{sec:chem}

%3
  \begin{figure*}
   \centering
  \includegraphics[width=17cm]{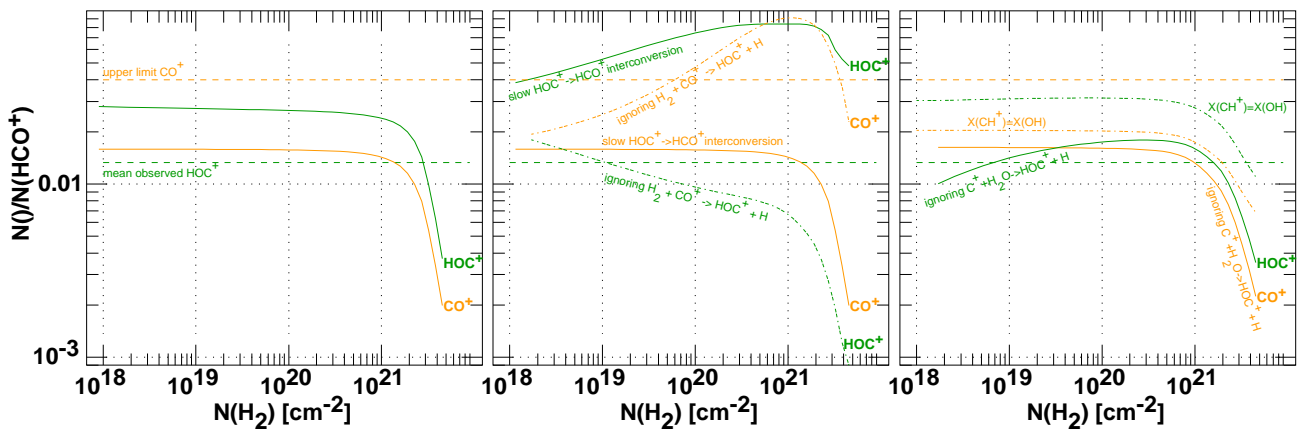}
      \caption{Models of the thermal chemistry for the relative abundances N(\cop)/N(\hcop) and N(\hocp)/N(\hcop)
 with default rates at left and deviations in the middle and right panels. In all panels results for \hocp\ are shown
 in  green and for \cop\ in orange. In all cases N(\hcop) is taken to be the observed value  
 N(\hcop) $= 3\times 10^{-9}$N(\HH).
  Left panel: Baseline models with the observed upper limit for N(\cop)/N(\hcop)  from this work and the mean 
 N(\hocp)/N(\hcop) ratio from \citet{gerin:19} illustrated schematically as horizontal dashed lines.  
 Middle panel: The solid lines represent the case that the rate constant of 
  \HH\ + \hocp\ $\rightarrow$ \hocp\ +\HH\ interconversion is taken as  $10^{-11}$\cccps.  
  The dashed-dotted lines represent model results when the rate constant for the reaction 
 \HH\ + \cop\ $\rightarrow$ \hocp\ + H is set to 0. 
Right panel: The solid lines show models in which the reaction of C\p\ and \HH O\ does not form \hocp\ and the dashed-dotted
 lines show model results when CH\p\ is as abundant as OH, X(CH\p) = $10^{-7}$. 
              }
       \label{fig:chem}
   \end{figure*}

Our upper limits  show that \cop\  cannot be responsible for forming the observed CO in 
diffuse molecular gas because \cop\ does not exist in sufficient quantity to replenish 
\hcop\ as it recombines with ambient electrons. However, beyond that, our limits combined 
with the observed relative abundance N(\hocp)/N(\hcop) 
$= 1.4\pm0.2 \times 10^{-4}$  \citep{gerin:19} can be used to gain further insights into the 
thermal chemistry that is working in diffuse molecular gas.  Most of the reactions that form 
\hcop\ also form \hocp, yet their observed abundances differ by a factor 70.  To explore this, 
we extended the thermal \cop\ and 
\hocp\  chemistry discussed by \cite{gerin:19} to include some hypotheticals corresponding to 
uncertainties in reaction rates in the KIDA reaction rate database \citep{Kida2014} and 
molecular abundances \citep{GerNeu+16}.  
As discussed in \citet{gerin:19}, some of the rates of the reactions controlling the 
formation of \cop\ and \hcop\ were not well known, especially the reaction 
$\rm{C^+ + OH \rightarrow CO^+ + H / CO + H^+}$ and the isomerization reaction destroying 
\hocp\ and forming \hcop, $\rm{HOC^+ + H_2 \rightarrow HCO^+ + H_2}$, whose rate is 
reported  as $4 \times 10^{-10}$\cccps\ at 25\, K and 300\, K \citep{smith:02}, 
about 1/5 the Langevin capture value and, in this temperature range, 
probably independent of the temperature.

As originally suggested by  \citet{GlaLan75,GlaLan76},
\cop\ can be formed by the reaction of C\p\ with OH, and also by the reaction of O with CH\p.  
\cop\ is destroyed by recombination with electrons and in chemical reactions with atomic 
and molecular hydrogen, and photo-dissociated with a free space rate $1\times10^{-10}$\ps\ 
that is always at least 300 times slower than the other destruction rates in the calculations 
discussed here. The OH relative abundance to \HH, X(OH) $= 1.0 \pm 0.14 \times10^{-7}$ 
varies only narrowly and is well-determined in local diffuse clouds \citep{WesGal+09,WesGal+10} 
such as those probed in this work. The excellent correspondence of the OH and \HH O absorption 
line profiles  and of the \HH O and \hcop\ profiles along long sight-lines across the Galactic 
plane \citep{wiesemeyer:16,gerin:19} further  indicates that this value of the OH abundance 
holds for the Galactic disk as well. 
The rate of the reaction of OH and C\p\ was recently calculated 
by  \citet{dagdigian:19}, providing accurate information on the rate coefficient and branching 
ratio over a temperature range 
10 -- 1000 \, K.  We approximate the rate constant for C\p\ + OH forming \cop\ as 
 $k =  2.2\times10^{-9}$(300/T)$^{0.107}$\cccps (see Table \ref{table:chem} ). By contrast 
the abundance of CH\p\ is locally variable and CH\p\ may not co-exist with the other species if it 
is predominantly formed in the non-thermal chemistry \citep{godard:14,valdivia:17}. We 
take X(CH\p) as a free parameter with X(CH\p) $ = 10^{-9}$ by default in the following discussion. 

\hocp\ can be formed in the reaction of \HH\ with \cop\ and in the reaction of  C\p\ with 
\HH O whose fixed relative abundance X(\HH O) $=2.4\times10^{-8}$ we take from \cite{GerNeu+16}.  
We ignore \hocp\ formation by the reaction of O with CH$_3$\p\ because the formation of CH$_3$\p\
is an aspect of the non-thermal chemistry.  \hocp\ recombines with electrons and is susceptible 
to photodissociation, but most importantly is isomerized to \hcop\ in reaction with \HH.  
The rate constant for the isomerization reaction of \HH\ and \hocp\ has alternatives 
in the KIDA database as discussed previously. Here we consider the two extreme values 
for the rate constant, $10^{-1  1}$\cccps\ given for 10-280 K and $4.7\times10^{-10}$\cccps\ given at 305 K.

The \cop\ and \hocp\ chemistry is simple when other molecular abundances are held fixed but it
involves the thermal and ionization balance for the temperature-sensitive recombination rates
and the atomic/molecular hydrogen balance for various reactions with \cop\ and \hocp.  
To treat this we performed calculations like those done earlier to account for the observations 
of HF and CF\p\ \citep{LisGuz+15}. In brief we calculated the self-consistent
global atomic/molecular, ionization and thermal equilibrium in gas spheres of constant 
total hydrogen number and column density, subject to the usual interstellar radiation fields. 
The hydrogen column density was varied and molecular abundances were integrated along the 
central line of sight to form the results that are displayed here. Plotting the results 
against N(\HH) removes some model sensitivities, for instance to the impact parameter 
about the cloud center used for the line of sight integration.

The results are shown in Fig. \ref{fig:chem}. At left is a baseline model using default values 
(Table \ref{table:chem}). The model produces \cop\ relative abundances about 2.5 times smaller than 
our observed upper limits, and about twice as much \hocp\ as observed. Declines in the abundances of 
\cop\ and \hocp\ for very high N(\HH) arise  from the recombination of the free gas-phase carbon 
from C\p\ to neutral carbon and CO.

In the middle panel the solid lines show results when \HH-catalyzed
\hocp\ $\rightarrow$ \hcop\ isomerization has the rate constant 
$10^{-11}$\cccps, a factor 2.5 above the default value.
The green curve for \hocp\ shows that the predicted abundance of \hocp\ grows even 
further above what is observed when the \hocp\ $\rightarrow$ \hcop\ interconversion is slow.
The orange curves show what happens when the rate constant for \hocp\ formation via 
\HH\ + \cop\ $\rightarrow$ \hocp\ + H is set to 0. The predicted amount of \cop\
grows well above our upper limits and the abundance of \hocp\ falls far below the observed mean.
We conclude that \hocp\ is forming mostly via C\p\ + OH followed by \cop\ + \HH\, but much or most of the \hocp\ so
produced is lost to isomerization into \hcop.

In the right side panel of Fig. \ref{fig:chem} the solid curves show model results when the reaction of C\p\ and
\HH O does not form \hocp:  all of the observed \hocp\ can indeed be made by the reaction of
C\p\ and OH, and agreement  with the observed \hocp/\hcop\ ratio is somewhat better when 
the contribution from \HH O is neglected. 
 However,  \HH O\ is observed to be present with 
X(\HH O) $=2.4\times10^{-8}$ \citep{GerNeu+16} and would be expected to form some
amount of \hocp. The default model may be predicting an overabundance of \cop\ 
or perhaps some aspect of the reaction of C\p\ and \HH O is not fully understood: a slower rate 
or a branching ratio that favors \hcop would lead to a lower contribution of this reaction to 
the \hocp\ formation.

CH\p\ may not co-exist with the other species if it is predominantly formed in the non-thermal 
chemistry \citep{godard:14,valdivia:17} and X(CH\p) is negligible ($10^{-9}$) in the default model.
The final deviation from the default model is represented by the dash-dot lines in the righthand
panel showing results when X(CH\p) = X(OH) = $10^{-7}$ is large.  The reaction of O + CH\p\ 
becomes an important source of \cop\ in the thermal chemistry when X(CH\p) $\ga 5\times10^{-8}$ 
and our upper limits on \cop\ by themselves imply X(CH\p)/X(OH) $\la 3$.  At such high CH\p\ 
abundances the discrepancy in \hocp\ grows even larger.  

%t4
\begin{table*}
\caption{CO isotopologues J=2-1 observations and column densities}             
\label{table:co}      
\centering          
\begin{tabular}{l c c c c c}     % 6 columns 
\hline\hline   

Source   & line & Flux& $\sigma_{l/c}$ & $\int \tau \rmv$ & N$^a$\\ 
         &      & Jy  &                & km s$^{-1}$  & $10^{14}\pcc$   \\
           \hline
J1717-3342 & \coth & 0.818  & 0.0024  & 0.260(0.004) & 6.3(0.4) \\
           & \coei & 0.822  & 0.0123  & $<$ 0.055    & $<1.3$   \\
J1744-3116  & \coth & 0.244  & 0.0073  &2.975(0.023)  & 72.0(0.6)\\ 
           & \coei &  0.244 & 0.0084  &0.102(0.018)  & 2.5(0.4)\\
     
\hline                                   %inserts single line
\end{tabular}
\\
$^a$Upper limit is 3$\sigma$.
\end{table*}

%t5
\begin{table*}
\caption{Rates and rate coefficients}             
\label{table:chem}      
\centering          
\begin{tabular}{l c c c }     % 4 columns 
\hline\hline   

Reaction & Symbol & Rate Coefficient & Reference \\
           &         & $\pccc$\ps& \\
           \hline
C\p\ + OH $\rightarrow$ CO\p\ + H & k$_1$ & $2.2\times10^{-9}$(300/T)$^{0.107}$ & \cite{dagdigian:19} \\
\cop\ + \HH\ $\rightarrow$ \hcop\ + H &  k$_2$ &        $7.5\times10^{-10}$ &  KIDA \\
C\p\ + H$_2$O $\rightarrow$ \hcop\ + H &  k$_3$ &      $2.1\times10^{-9}$ &  KIDA \\
C\p\ + H$_2$O $\rightarrow$ \hocp\ + H &  k$_4$ &     $2.1\times10^{-9}$ &  KIDA, \cite{martinez:08} \\
\hocp\ + H$_2$ $\rightarrow$ HCO\p\ + H$_2$ &  k$_5$  & $4 \times 10^{-10}$ & \cite{smith:02} \\
\hocp\ + H$_2$ $\rightarrow$ HCO\p\ + H$_2$ & k$_6$  & $1 \times 10^{-11}$  & KIDA \\
\hcop\ + e$^-$  $\rightarrow$ CO + H & $\alpha(T)$& $1.43\times10^{-5}/{\rm T}^{0.69}$ & \cite{hamberg:14}\\
\hline
\hline
Reaction & Symbol & Rate  & Reference \\
         &      & s$^{-1}$ & \\
               \hline
CO + $h\nu$ $\rightarrow$ C + O  &  k$_d$ &$2.43 \times 10^{-10} {\rm exp}(-3.88 \Av) $ & KIDA, \cite{HeaBos+17}\\
CO$^+$ + $h\nu$ $\rightarrow$ C$^+$ + O  &   &$1.04 \times 10^{-10}  $ & KIDA\\
     
\hline                                   %inserts single line
\end{tabular}
\end{table*}

%4 
   \begin{figure}
   \centering
   \includegraphics[width=9cm]{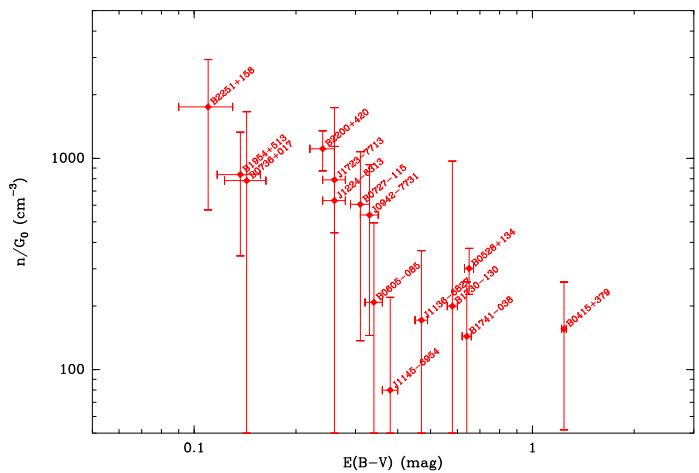}
   \caption{Number density $n({\rm H})/G$ vs reddening \EBV\ \citep{GreGre+19} derived by 
   equating CO formation and photo-destruction for sight-lines with \hcop\ and CO 
   column densities measured in absorption \citep{LisLuc98,LisGer+19}, as discussed 
   in Section 5.}
\label{fig:co}
   \end{figure}

\section{Conditions for the growth of CO}
\label{sec:co}

Balancing the formation of CO through dissociative recombination of \hcop\ and 
destruction of CO by photodissociation can constrain the physical conditions where CO 
might form in diffuse molecular gas.  With a free-space photodissociation
rate  $k_d = 2.43 \times 10^{-10}$ s$^{-1}$ for the Draine radiation field 
(defined as $G_0=1$) and
an attenuation  scaling as $e^{-\gamma \Av}$ this  balance locally implies

$$ \frac{n({\rm CO})}{n(\hcop)} = \frac{\alpha(T)x(e)n({\rm H})}{G_0k_d e^{-\gamma \Av}}  $$

For a uniform slab, integrating the attenuation term from 0 up to half the slab extinction, 

$$ \frac{N({\rm CO})}{N(\hcop)} = \frac{\alpha(T)x(e)n({\rm H})} {G_0k_d } \frac{0.5 \gamma  \Av}{ 1-e^{-0.5\gamma  \Av}}  $$

$n({\rm H})/G_0$ can therefore be expressed as

$$ n({\rm H})/G_0 = \frac{N({\rm CO})}{N(\hcop)}  \frac{k_d (1-e^{-0.5\gamma  \Av})}{\alpha(T)x(e)0.5 \gamma  \Av} $$

where we approximate \Av$ = 3.1$\EBV\ and take \EBV\ from \citet{GreGre+19}, replacing local
quantities with line of sight averages. The error bar in \EBV\ is a nominal value 0.02 mag.
We take $\gamma = 3.88$ and use the reaction rates
and rate coefficients given in Table \ref{table:chem}.  Fig. \ref{fig:co} presents the 
variation of $n({\rm H})/G_0$ as a function of the reddening \EBV\ for the existing sample of QSO sight 
lines with \hcop\ and CO column densities determined in absorption \citep{LisLuc98,LisGer+19}.
We use conditions as described in Section 2, a uniform temperature of 40K and a uniform electron 
fraction $x(e) = 2 \times 10^{-4}$.

The mean value for the sample is $<n({\rm H})/G_0 > \sim 300\pccc$.
For a standard interstellar radiation field of $G_0 = 1$, the  calculated densities n(H) $\sim 100-300 \pccc$ are modest for 
\EBV\ $\ga 0.5$ mag, even in the absence of any consideration of CO shielding 
by itself or by \HH. The higher densities $n({\rm H}) \sim 1000-2000\pccc$ calculated at
\EBV\ $\la$ 0.2 mag are higher by a factor $\sim$ 5 than are inferred from 
the brightness of \hcop\ emission in these directions \citep{Lis20Hotspots}.  Thermal 
pressures  $p/k \sim$ 40 K  $\times$ 1000$\pccc$  are an order of magnitude higher 
than typical values in the diffuse molecular gas \citep{JenTri11} and would more 
likely represent the conditions under which a turbulence-driven chemistry forms 
\hcop. A lower value of the radiation field toward these sight-lines would bring the derived densities closer to the mean values.
The data shown in this plot include the sight lines with CO and \hcop\ detections,  which may bias the derived $n({\rm H}/G_0$ to
somewhat high values as a higher density implies  a faster CO production rate.

The inverse variation of $n({\rm H})/G_0$ with \EBV\ in Fig. \ref{fig:co}, mirroring the 
functional dependence on Av, reflects the fact that the observed values of 
N(CO)/N(\hcop) are not correlated with \EBV\ and show a moderate scatter of at most a factor of 1. A  majority of the selected  sight-lines have
N(CO)/N(\hcop) between 1000 and 2000.  We also plotted
$n({\rm H})/G_0$ against the molecular fraction \fH2\ = 2N(\HH)/N(H) taking N(\HH) = 
N(\hcop)/$3\times10^{-9}$ and N(H) =  $a \times$ \EBV.  Requiring \fH2\ $\leq 1$
in all cases implies $a \ga 8\times 10^{21}~{\rm H-nuclei}\pcc$ mag$^{-1}$ but 
there is no trend for $n({\rm H})/G0$ to vary with \fH2\ calculated in this way. 

\section{Summary and conclusions}
\label{sec:conclusion}

We sought to test the origin of CO in diffuse molecular gas within the generally
accepted framework that CO is predominantly formed through the dissociative electron 
recombination of \hcop. \hcop\ is widely observed in diffuse molecular gas, with
a relative abundance N(\hcop)/N(\HH) $= 3\times 10^{-9}$. 

In a quiescent, thermal, cosmic-ray and UV-driven chemistry, \hcop\ is predominantly formed 
in the reactions C\p\ + OH $\rightarrow$ \cop\ + H and \cop\ + \HH$ \rightarrow$ \hcop\ + H.
Using Cycle 7 ALMA Band 6 observations at 236 GHz (Table 1) we searched for \cop\ absorption 
toward two bright compact extragalactic mm-wave continuum sources seen at low latitude 
in the inner Galaxy and known to be occulted by high column densities of \hcop-bearing 
diffuse molecular gas (see Section 2 and Table 2).  We failed to detect \cop\ at levels 
sufficient to demonstrate in Section 3 that the reaction \cop\ + \HH\ $\rightarrow$ \hcop\ 
+ H cannot replenish more than a few percent of the observed \hcop\ that recombines.
The non-detection of \cop\  confirms that \hcop\ is mainly produced in the reaction 
between oxygen and carbon hydrides CH$_2^+$ or CH$_3^+$ induced by supra-thermal 
processes while \cop\ and \hocp\  result from reactions between C$^+$ and OH and 
\HH O\ in the thermal chemistry that occurs in quiescent diffuse 
molecular gas. The observed \cop\ and \hocp\ abundances relative to \HH\ set an upper limit 
on the rate coefficient of these reactions.

In Section 4 we  explored the coupled thermal chemistries of \cop\ and \hocp\ given
our observational upper limits on \cop\ and existing observations of \hocp\ 
showing N(\hocp)/N(\hcop) $= 0.015\pm0.003$.  A baseline cosmic-ray and UV-driven thermal
chemical model overproduced \hocp\ by a factor two, but by as much as a factor
six if the isomerization reaction \hocp\ + \HH\ $\rightarrow$ \hcop\ + \HH\ does
proceed rapidly. The option of a low rate for this reaction in the KIDA database
should probably be ignored.  The predicted \hocp\ abundance is in better
agreement with observation if the reaction of C\p\ and \HH O preferentially forms
\hcop, rather than \hocp.  We placed a weak limit on the relative abundance
of CH\p\ in quiescent gas, N(CH\p)/N(OH) $\la$ 3 or X(CH\p)/X(\HH) $ \la 3\times10^{-7}$
such that the reaction O + CH\p\ does not become an important source of \cop,
substantially over-producing it. 

In Section 5 we derived characteristic number densities $n({\rm H})$ at which CO would form
in diffuse molecular gas with observed abundances.  For \EBV\ $\ga$ 0.5 mag the
densities are modest, $n({\rm H}) \sim 100-300\pccc$ but for CO-bright regions of low 
extinction \EBV\ $ \sim 0.1-0.2$ mag they are an order of magnitude
higher $n({\rm H}) \sim 1-2\times10^3 \pccc$ and such heavily over-pressured gas 
is more properly considered in the context of a turbulent dissipation chemistry. The presence of CO in sight-lines of low reddening could also indicate a lower radiation field intensity.

\begin{acknowledgements}
The authors thanks the referee Helmut Wiesemeyer for his thoughtful comments and suggestions. This paper makes use of the following ALMA data: ADS/JAO.ALMA\#2019.1.00120.S.. ALMA is a partnership of ESO (representing its member states), NSF (USA) and NINS (Japan), together with NRC (Canada), NSC and ASIAA (Taiwan), and KASI (Republic of Korea), in cooperation with the Republic of Chile. The Joint ALMA Observatory is operated by ESO, AUI/NRAO and NAOJ. The National Radio Astronomy Observatory is a facility of the
National Science Foundation operated under cooperative agreement by Associated Universities, Inc. This work was supported by the Programme National ''Physique et
Chimie du Milieu Interstellaire'' (PCMI) of CNRS/INSU with INC/INP co-funded
by CEA and CNES. 
\end{acknowledgements}

\bibliographystyle{aa}
%\bibliography{coplus,absorption,mnemonic}

\end{document}